\documentclass[10pt,journal]{IEEEtran}

\usepackage{cite}
\usepackage[pdftex]{graphicx}
\usepackage{ulem}
\usepackage{xcolor}
\usepackage{siunitx}

\usepackage{array}
\newcolumntype{P}[1]{>{\centering\arraybackslash}p{#1}}

\begin{document}

\title{Fabrication Development for SPT-SLIM, a Superconducting Spectrometer for Line Intensity Mapping}

\author{%
T. Cecil, C. Albert, A.~J.~Anderson, P.~S.~Barry, B. Benson, C. Cotter, C. Chang, M. Dobbs, K. Dibert, R. Gualtieri, K.~S.~Karkare, M. Lisovenko, D. P. Marrone, J. Montgomery, Z. Pan, G. Robson, M. Rouble, E. Shirokoff, G. Smecher, G. Wang, and V. Yefremenko% <-this % stops a space

\thanks{Manuscript receipt and acceptance dates will be inserted here. Work at Argonne, including use of the Center for Nanoscale Materials, an Office of Science user facility, was supported by the U.S. Department of Energy, Office of Science, Office of Basic Energy Sciences and Office of High Energy Physics, under Contract No. DE-AC02-06CH11357. This work was supported by Fermilab under award LDRD-2021-048 and by the National Science Foundation under award AST-2108763. The McGill authors acknowledge funding from the Natural Sciences and Engineering Research Council of Canada and Canadian Institute for Advanced Research. 

(\textit{Corresponding author: Thomas Cecil)}}% <-this % stops a space
\thanks{T. Cecil is with the Argonne National Laboratory, Argonne, IL 60439 USA (email: cecil@anl.gov)}%
\thanks{C. Albert was with the the University of Chicago, Chicago, IL 60637 USA and is now with Caltech, 1200E California Blvd, Pasedena CA 91125 USA (email: calbert@caltech.edu)}
\thanks{A.~J.~Anderson, B. Benson, and K. Karkare are with Fermi National Accelerator Laboratory, PO BOX 500, Batavia, IL 60510 and the University of Chicago, Chicago, IL 60637 USA (email: adama@fnal.gov; kkarkare@kicp.uchicago.edu; bbenson@astro.uchicago.edu)}
\thanks{P. Barry and G. Robson are with Cardiff University, Cardiff CF10 3AT, UK (email: barryp2@cardiff.ac.uk; RobsonG2@cardiff.ac.uk)}
\thanks{C. Cotter, K. Dibert, and E. Shirokoff are with the University of Chicago, Chicago, IL 60637 USA (email:  cotter@astro.uchicago.edu; krdibert@uchicago.edu; shiro@uchicago.edu) }
\thanks{C. Chang is with the Argonne National Laboratory, Argonne, IL 60439 USA and the University of Chicago, Chicago, IL 60637 USA (email: clchang@kicp.uchicago.edu) }%
\thanks{M. Dobbs is with McGill University, 845 Rue Sherbrooke O, Montreal, QC H3A 0G4, Canada and the Canadian Institute for Advanced Research, Toronto, ON, M5G 1Z8, Canada  (email: Matt.Dobbs@mcgill.ca)}
\thanks{R. Gualtieri, M. Lisovenko, Z. Pan, G. Wang, and V. Yefremenko, are with the Argonne National Laboratory, Argonne, IL 60439 USA (email: rgualtieri@anl.gov, mlisovenko@anl.gov; panz@anl.gov, gwang@anl.gov; yefremenko@anl.gov)}%
\thanks{D. P. Marrone is with the University of Arizona, Tucson, AZ 85721 (email: dmarrone@email.arizona.edu)}
\thanks{J. Montgomery, and M. Rouble are with McGill University, 845 Rue Sherbrooke O, Montreal, QC H3A 0G4, Canada (email: joshua.j.montgomery@gmail.com; maclean.rouble@mcgillcosmology.ca)}
\thanks{G. Smecher is with Three Speed Logic, Inc., Victoria, BC, Canada (email: gsmecher@threespeedlogic.com)}
}

% The paper headers
\markboth{ASC2022-2EPo2E-01}%
{Shell \MakeLowercase{}}

\IEEEtitleabstractindextext{%
\begin{abstract}
Line Intensity Mapping (LIM) is a new observational technique that uses low-resolution observations of line emission to efficiently trace the large-scale structure of the Universe out to high redshift. Common mm/sub-mm emission lines are accessible from ground-based observatories, and the requirements on the detectors for LIM at mm-wavelengths are well matched to the capabilities of large-format arrays of superconducting sensors. We describe the development of an $R$ = $\lambda / \Delta\lambda = 300$ on-chip superconducting filter-bank spectrometer covering the 120--180 GHz band %optimized
for future mm-LIM experiments, focusing on SPT-SLIM, a pathfinder LIM instrument for the South Pole Telescope. Radiation is coupled from the telescope optical system to the spectrometer chip via an array of feedhorn-coupled orthomode transducers. Superconducting microstrip transmission lines then carry the signal to an array of channelizing half-wavelength resonators, and the output of each spectral channel is sensed by a lumped element kinetic inductance detector (leKID). Key areas of development include incorporating new low-loss dielectrics to improve both the achievable spectral resolution and optical efficiency and development of a robust fabrication process to create a galvanic connection between ultra-pure superconducting thin-films to realize multi-material (hybrid) leKIDs. We provide an overview of the spectrometer design, fabrication process, and prototype devices.

\end{abstract}

}

% make the title area
\maketitle

\IEEEdisplaynontitleabstractindextext

\IEEEpeerreviewmaketitle

\section{Introduction}
\label{sec:introduction}

\IEEEPARstart{S}{tudies} of the mm-wave sky have provided a wealth of information about the history of the universe. The cosmic microwave background (CMB) at a redshift of $z \sim 1100$ provides a window to the earliest stages of the universe as high-energy photons are redshifted down into the mm-wave band. Similarly, far-infrared photons in the later universe are redshifted into the mm-wave band. Line intensity mapping (LIM) is a technique that uses integrated line emissions to trace the large-scale structure of the Universe out to high redshift \cite{kovetz2017}. LIM techniques have the potential to constrain cosmology (e.g., inflation, dark energy, neutrino mass) with higher precision than CMB and galaxy surveys  \cite{moradinezhad2019, karkare2018, moradinezhad2022}. Many common mm/sub-mm emission lines are accessible from ground-based observatories, matching with atmospheric transmission windows. Additionally, the requirements for detectors in future LIM experiments are well aligned with existing capabilities achievable with large-format arrays of superconducting sensors \cite{Karkare.2021}. Many early detector plans for LIM experiments \cite{Endo.2012,Shirokoff.2014,cataldo_2019} draw on the heritage from decades of CMB and leKID detector development incorporating features such as low-loss superconducting microstrips \cite{Duff.2016,Posada.2015} , othrho-mode transducer (OMT) optical coupling \cite{Henderson.2016,Galitzki.2018}, and leKIDs \cite{Doyle.2008} to realize OMT coupled KIDs\cite{Brien.2018,Lourie.2018,Tang.2020}. The combination of accessible mm/sub-mm emission lines and large-format arrays of superconducting sensors makes LIM experiments a promising pathway for future cosmological breakthroughs. 

The spectrometer development described below is aimed at producing the focal plane for the planned South Pole Telescope Summertime Line Intensity Mapper (SPT-SLIM) experiment \cite{Karkare.2022}. SPT-SLIM is a pathfinder LIM instrument for the South Pole Telescope. It is designed to use the same primary mirror as the SPT-3G experiment with new pickoff mirrors and a dedicated cryostat. The cryostat will be cooled to $\sim$100 mK through the combination of a pulse tube cryocooler and an adiabatic demagnetization refrigerator (ADR). 
%SPT-SLIM will have a field of view of XXX and a usable focal plan area of XXX.
The focal plane will be filled with 18 dual-polarization pixels, each coupled to an $R$ = $\lambda / \Delta\lambda = 300$ microstrip spectrometer covering the 120--180 GHz band. Additional details on the spectrometer design  are provided below in Section \ref{SpecDesignSec}. SPT-SLIM will be the first demonstration of on-chip spectroscopy in LIM-style observations that is sensitive to CO emission from $0.5 < z < 3$. SPT-SLIM is scheduled to deploy to the South Pole for observations in the 2023--2024 austral summer. 

\section{Spectrometer design}
\label{SpecDesignSec}
\begin{figure}
    \centering
    \includegraphics[width=1.0\linewidth]{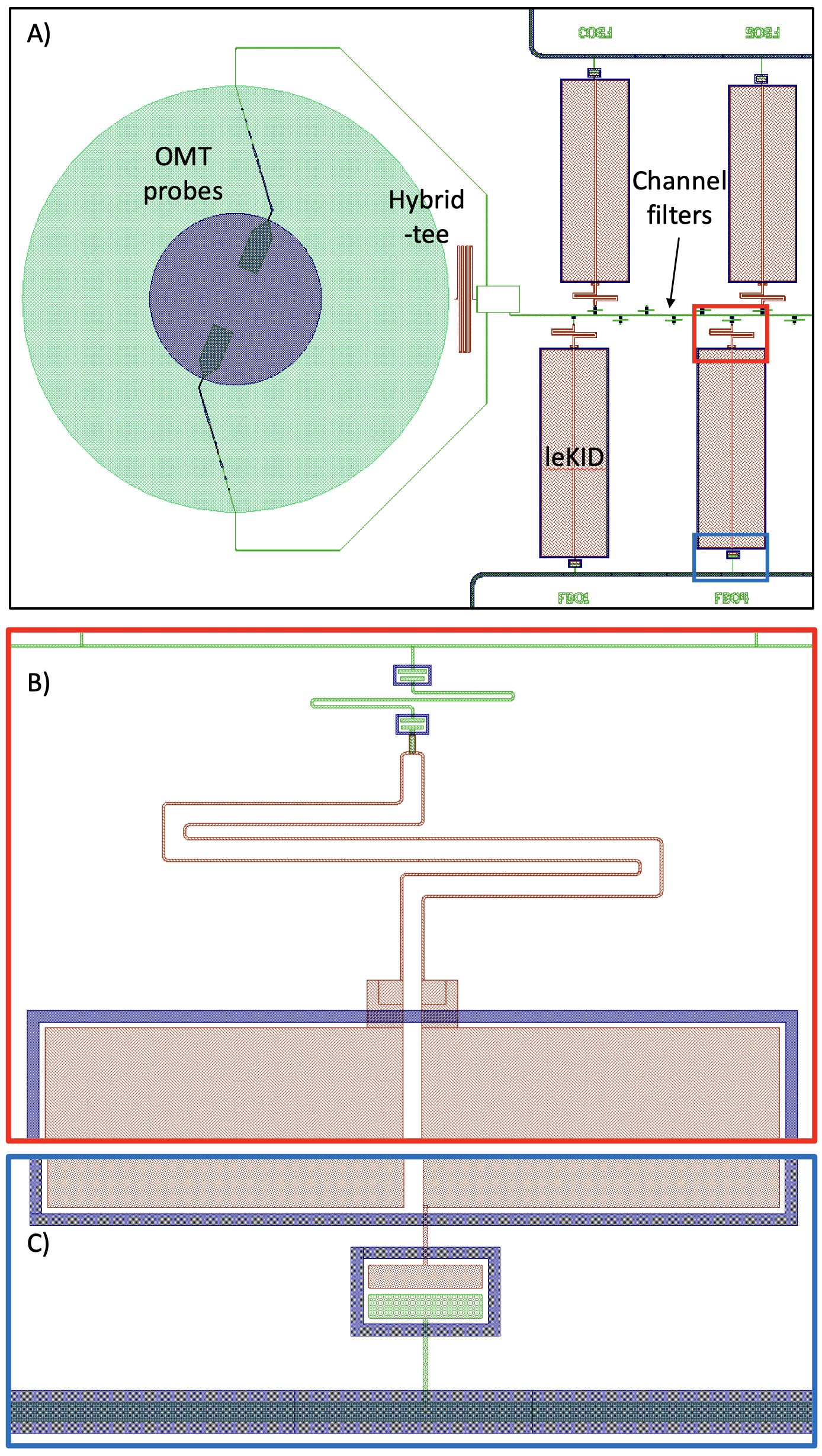}
    \caption{Layout for a filter bank spectrometer. (A) Layout showing main spectrometer features including the OMT probes, hybrid-tee, central microstrip, channelizing filters, and readout leKIDs. (Note: while SPT-SLIM will use dual-polarization pixels we show a single polarization for simplicity) (B) Zoom-in showing channelizing filter coupled to the meander inductor and capacitor of readout leKID. (C) Zoom-in showing readout leKID connected to the readout microstrip via a coupling capacitor. For a description of the entire focal plane the reader is directed to \cite{Barry.2021zfa}.}
    \label{fig:SpecDesign}
\end{figure}

SPT-SLIM employs filterbank-style spectrometers \cite{Barry.2021zfa,Robson.2021}. The key components for the spectrometer are shown in Figure \ref{fig:SpecDesign}. Optical signals from the telescope are coupled to the spectrometer via a feedhorn to an ortho-mode transducer (OMT) coupling. The signal from opposing OMT probes are combined in a hybrid-tee and the resulting signal continues down a microstrip transmission line. For each filter bank channel one end of a $\lambda/2$ channelizing resonator is capacitively coupled to the microstrip feedline, with each filter picking off a narrow frequency band. The channel filters are arranged from high to low frequency along the feedline with the higher-frequency channels closer to the hybrid-tee. The channel filters are designed to oversample the 120--180 GHz band. The other end of the channelizing resonator is  capacitively coupled to the inductor of a lumped element kinetic inductance detector (leKID) that is used to detect the power in each channel and provides readout multiplexing. The capacitor of the leKID is capacitively coupled to a co-planar waveguide that serves as the readout line. The leKIDs are designed to resonate at $\sim 2$ GHz with a frequency spacing of 500 kHz between each resonance. The frequency range was selected to match the chosen readout hardware \cite{Rouble.2022} and balance the trade-off between lower resonance frequencies and larger resonator size.%This will enable placing 200 channels in a 500 MHz readout bandwidth.

%%%%%%%%%%%%%%%%%%%%%%%%%%%%%%%%%%%%%%%%%%%%%%%
\section{Fabrication process}

\begin{figure}
    \centering
    \includegraphics[width=1.0\linewidth]{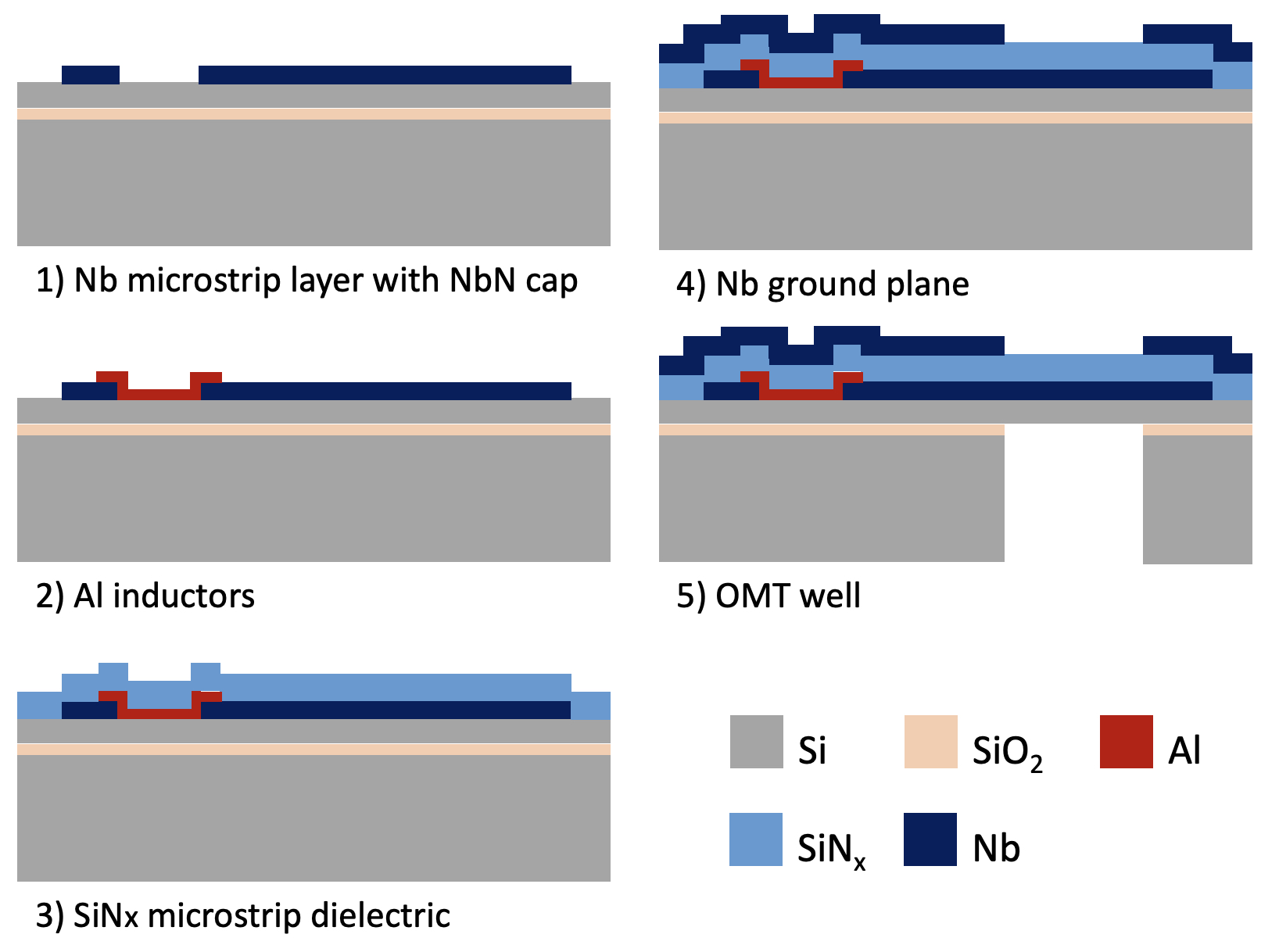}
    \caption{Process flow for fabrication of a superconducting spectrometer with inverted microstrip on a SOI wafer.}
    \label{fig:ProcessFig}
\end{figure}

Spectrometer fabrication was carried out at the Center for Nanoscale Materials at Argonne National Laboratory. The superconducting microstrip and capacitors of the readout leKID are made from a Nb-NbN/dielectric/Nb stack that uses a silicon-based dielectric as described in Section \ref{DielectricsSec}. The readout leKID is a hybrid device that pairs the Nb capacitors to an Al inductor. A NbN cap layer is grown on the top of the bottom Nb layer to provide a galvanic contact to the Al inductor as described in Section \ref{NbNSection}.  Nb was used in the capacitors both due to reports of lower two-level systems (TLS) noise levels for Nb grown on Si as compared to Al on Si \cite{Zmuidzinas.2012} and to boost the effective kinetic inductance fraction by minimizing additional kinetic inductance from the capacitor.

The process flow to fabricate the filter bank spectrometer is shown in Figure \ref{fig:ProcessFig}. The spectrometer begins with a silicon on insulator wafer (SOI) with a high-resistivity device layer ($>10$ kOhm-cm) to limit losses from the substrate. The spectrometer design uses an inverted microstrip layout; using this geometry reduces the number of processing steps prior to the deposition of the Al inductors and deposits the Al layer on the high-resistivity silicon substrate in an effort to yield the highest quality Al film. The first layer is the spectrometer wiring including OMT probes, transmission lines, channelizing filters, and leKID capacitors. This layer consists of 170 nm thick Nb with a 30 nm thick NbN capping layer, which is used to form a galvanic contact to the Al inductor (for additional details on the NbN capping layer see Section \ref{NbNSection}). The Nb/NbN layer is patterned using a fluorine-based reactive ion etch. Next the 30 nm Al inductors are patterned using a bilayer liftoff process. The Al thickness is chosen based upon the inductor volume  to achieve the required responsivity for the expected optical loading from the South Pole \cite{Barry.2021zfa}. Following the wiring layers is the dielectric layer. The current baseline is a Si-rich SiN$_x$ deposited by plasma-enhanced chemical vapor deposition (PECVD) at a temperature of 100$^{\circ}$C to avoid damaging the Al inductors. After the dielectric, the Nb ground plane is deposited and patterned via liftoff. Using a liftoff process avoids any shorting that could occur from material that is deposited on the sidewall of steps; sidewall material may not be removed by an anisotropic reactive ion etch process. Next, the OMT membrane is formed by a deep reactive ion etch (DRIE) from the back side of the wafer to remove the SOI handle wafer from behind the OMT probes. Lastly, a wet etch with 10$\%$ HF buffered oxide etch is used to remove the buried oxide layer in the OMT well. %Finally a layer of Nb is deposited on the backside to create a quarter-wave backshort. The same lithography layer that is used for the DRIE step also serves as a liftoff layer for the Nb backshort.

Realization of this fabrication process required addressing key technical challenges which are addressed in more detail below. The first key challenge was identifying a low-temperature dielectric deposition process that yields a low-loss dielectric. The second challenge was achieving a robust galvanic contact between the Al and Nb layers in the hybrid leKIDS despite the presence of native oxides. The last challenge seeks an alternative substrate to SOI wafers. Although high quality leKIDs have been fabricated on SOI \cite{Vissers.2020} the increased cost and lead time for SOI wafers with high-resistivity device layers makes them difficult to work with in a development project where detector designs are still iterating. We explore an alternate substrate using isolated SiN$_x$ membranes.

\subsection {Low-loss Dielectrics}
\label{DielectricsSec}
The quality of the dielectric used in the microstrip transmission lines and half wave channelizing filters plays a key role in the overall performance of the spectrometers. The dielectric loss tangent sets a practical limit on the spectrometer's resolution by placing a lower limit on the channelizing filter bandwidth. It also impacts the overall filter efficiency. Previous work \cite{Robson.2021} simulating the performance of the SPT-SLIM filters targets a loss tangent of $7\times10^{-4}$ at mm-wave frequencies. From microstrip dielectric loss measurements we expect the loss tangent at mm-wave to be up to a factor of 10 higher than the loss tangent measured at microwave frequencies \cite{Hähnle.2021} and thus seek a dielectric with a loss tangent near $1\times10^{-4}$ at microwave frequencies. A wide range of potential dielectrics are available with many of these achieving the desired loss tangent, e.g., SiN$_x$ \cite{Paik.2010,Duff.2016}, amorphous silicon (a-Si) \cite{Hähnle.2021} and amorphous silicon carbide (a-SiC) \cite{Buijtendorp.2021}. However, many of the deposition processes used for these high quality dielectrics involve heating the substrate to temperatures of several hundred degrees ($\sim 250^{\circ}$C to 350$^{\circ}$C). This poses challenges for the inverted microstrip process. We have observed degradation to Al inductors - formation of hillocks - when heated above $\sim150^{\circ}$C and avoid any subsequent processing temperatures above this. We surveyed a range of different materials and deposition processes to identify  a low-loss dielectric that can be deposited at low temperatures \cite{Lisovenko.2022}. A summary of materials and processes used is shown in Table 1. The ideal dielectric will have low intrinsic stress, high thickness uniformity, and limited pin holes. a-Si deposited via chemical vapour deposition (CVD) meets many of these criteria, but requires substrate temperatures above 250$^{\circ}$C. Alternatives process with lower temperature requirements include PECVD and ion-beam assisted sputtering (IBAS). The IBAS process is especially attractive as it allows for room temperature depositions compatible with liftoff processing. After evaluating several films we have identified a Si-rich SiN$_x$ film deposited via PECVD that meets all the criteria for use with inverted microstrip. With a loss tangent of $2\times10^{-4}$ at microwave frequencies ($\sim$1 GHz) this dielectric loss is the lowest of the evaluated dielectrics with a low-temperature process. Additional testing of the loss tangent at mm-wave frequencies will be needed to make sure the performance does not degrade at higher frequencies.

\begin{table}
\caption{\label{DielectricTable} Dielectric materials characterized in the search for a low temperature deposition process for low-loss dielectrics}

\begin{tabular}{ P{1.75cm} P{1.75cm} P{1.75cm} P{1.75cm}  }
 \hline
 \hline
Material & Process & Temp & Loss Tangent \\
&&(C) & ($\sim$ 4 GHz) \\
 \hline
 SiN   & CVD    & 275 &  2E-3 \\
 a-Si & CVD  & 275   & 4E-5 \\
 SiN & PECVD &  100 & 2E-3 \\
 Si-rich SiN & PECVD & 100 & 2E-4\\
 SiO$_2$ & IBAS & 20 & 8E-4\\
 SiN & IBAS & 20  & 2E-3\\
 Si-rich SiN & IBAS & 20 & 8E-4\\
 a-Si & IBAS & 20  & 5E-4\\
 \hline
\end{tabular}
% \caption{\label{DielectricTable} Dielectric materials characterized in the search for a low temperature deposition process for low-loss dielectrics}
\end{table}

\subsection {Galvanic Contacts Using a NbN Interface Layer}
\label{NbNSection}
\begin{figure}
    \centering
    \includegraphics[width=1.0\linewidth]{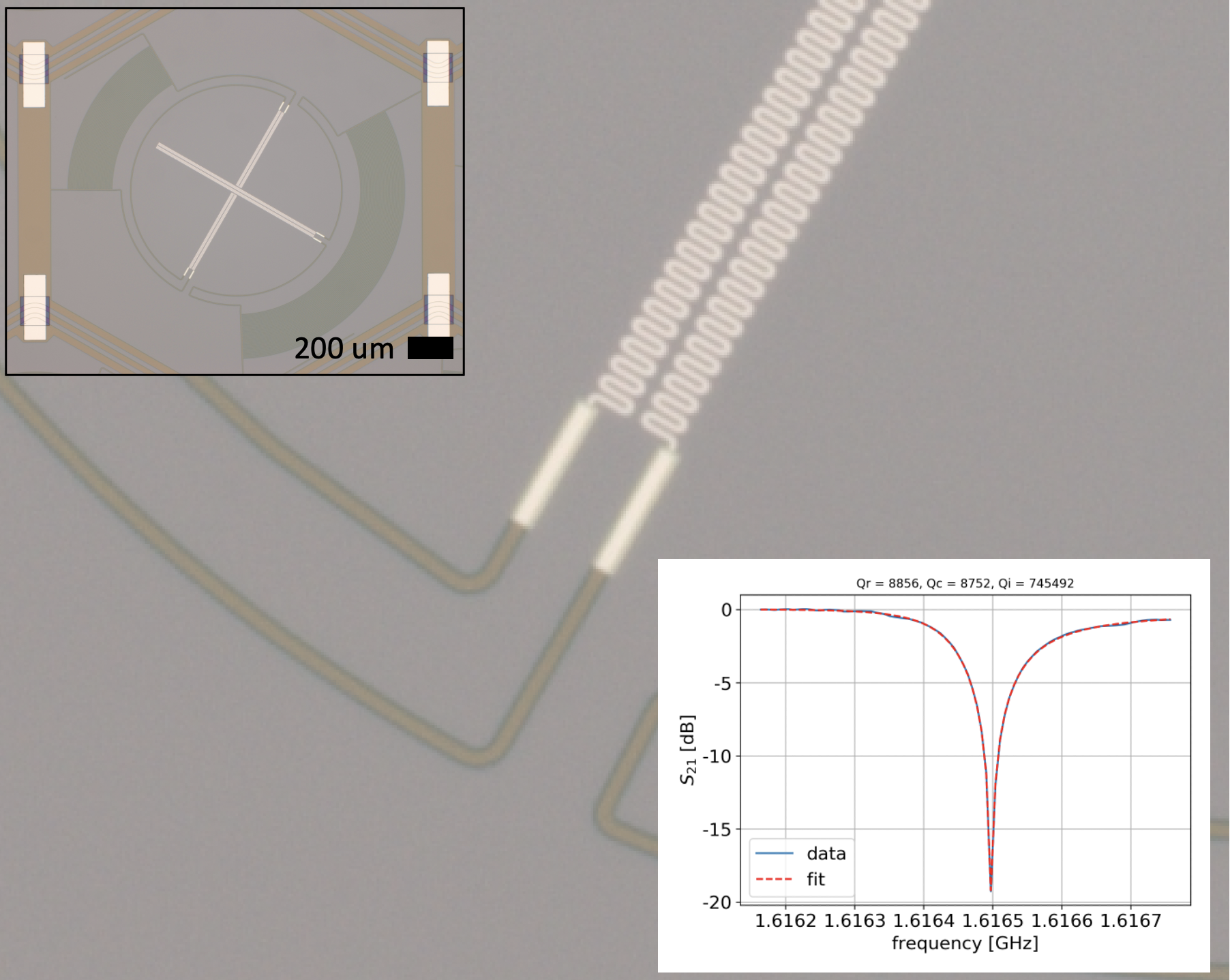}
    \caption{Photo of a SPT-3G+ style hybrid leKID with Al inductor and Nb capacitor used to verify the RF performance of the NbN capping layer. A 30nm layer of NbN caps the Nb and provides galvanic contact between the Nb and Al. (Upper left inset) Photo of complete leKID showing meander inductor and interdigitated capacitor. (Lower right inset) Plot of S21 frequency sweep of leKID with corresponding parameter fit, indicating good performance of the NbN cap layer due to the high leKID quality factor.}
    \label{fig:AlNbFig}
\end{figure}

The hybrid structure of the kinetic inductance detectors---Al inductors and Nb capacitors---poses an additional fabrication challenge in ensuring a clean galvanic contact between the Al and Nb layers. Early attempts at fabricating inverted microstrip devices deposited the Al inductors first followed by the Nb capactors. These devices showed inconsistent contact between the layers with some devices open and others with various level of contact resistance.
%Devices for SPT-SLIM will be fabricated at both the University of Chicago and Argonne; 
The deposition system at Argonne does not have an ion mill, but historically has used an RF substrate bias to clean surfaces prior to deposition. This cleaning step proved insufficient in removing the native aluminum oxide from the inductors prior to the subsequent Nb deposition. Similar devices fabricated at the University of Chicago using an ion mill were sensitive to the ion mill settings and also yielded inconsistent contact between the layers. To provide a clean galvanic contact between the layers, the process steps were flipped with the Nb deposited first and a thin ($\sim30 $~nm) NbN capping layer added to the top of the Nb. The NbN cap limits oxidation of the Nb when the Nb layer is removed from vacuum and patterned. Both DC and RF test structures have been fabricated to evaluate the performance of the cap layer. The DC test structures consist of a pair of Nb-NbN-Al-NbN-Nb connections with various contact geometries. The measured resistance of the DC test structures dropped to zero when below the Al transition temperature with no sign of an interface resistance. To verify the performance at microwave frequencies we fabricated leKIDs using an SPT-3G+ design we have previously characterized and shown to have high quality factors \cite{Dibert.2022,Pan.2022}. A photo of one such device along with a trace of $|S_{21}|^2$ for an example resonator are shown in Figure \ref{fig:AlNbFig}. Hybrid leKIDS made using the NbN capping layer demonstrated high internal quality factors, indicating the NbN layer was not degrading device performance.

\subsection {Membrane Step Down}
\begin{figure}
    \centering
    \includegraphics[width=1.0\linewidth]{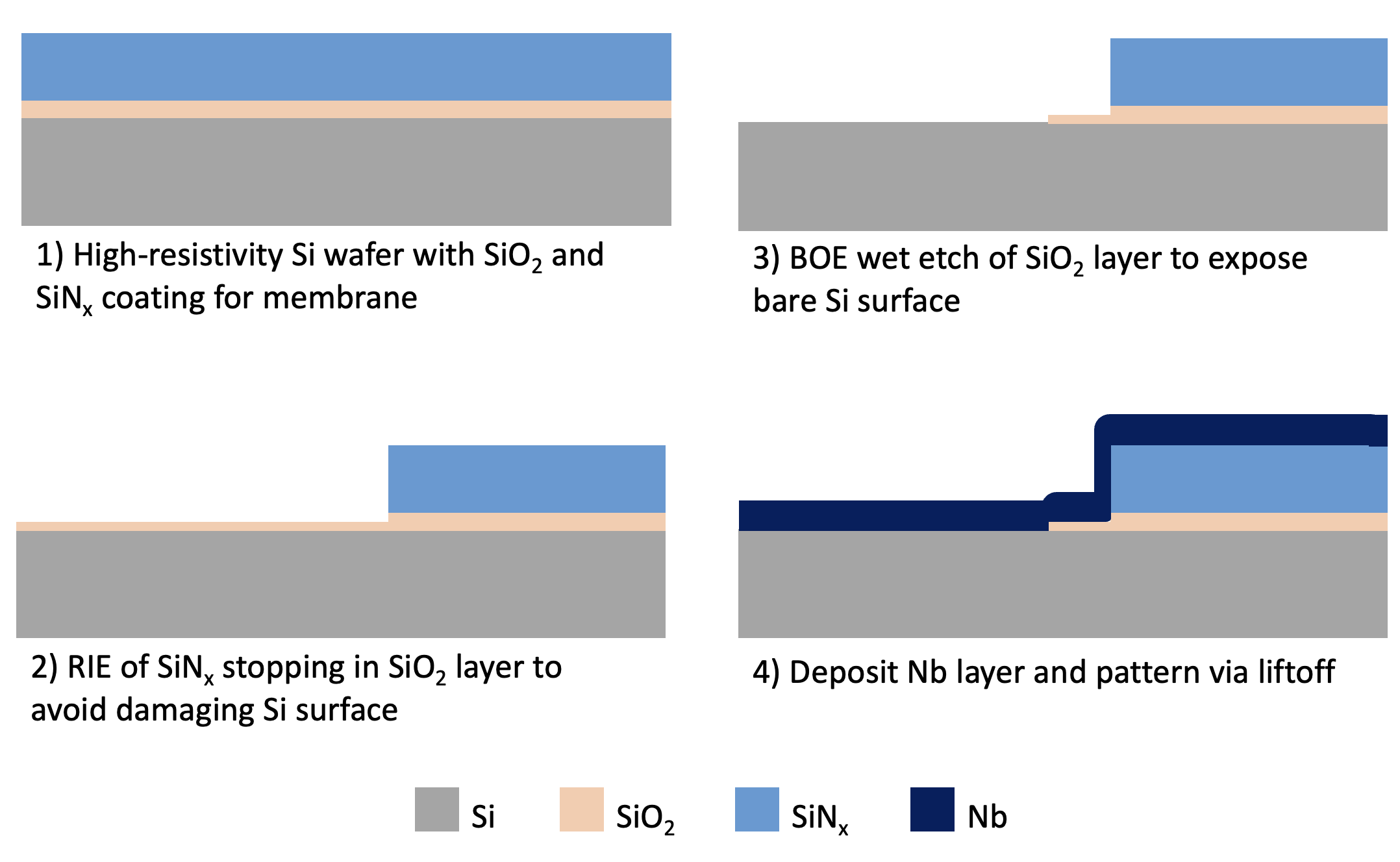}
    \caption{Fabrication steps for membrane step down process. The SiO$_2$ layer serves as an etch stop layer for the SiN$_x$ etch to avoid damaging the Si surface.}
    \label{fig:MembraneFig}
\end{figure}

The baseline design for SPT-SLIM calls for the spectrometers to be fabricated on a silicon on insulator (SOI) wafer with a high-resistivity device layer. This provides a high-quality surface on which to deposit the inverted microstrip layers. However, SOI wafers can be significantly more expensive than bulk wafers and can have long lead times when requiring specific device layer properties (e.g., thickness and resistivity). As an alternative we are exploring using bulk high-resistivity Si wafers coated with thin SiO$_2$ and SiN$_x$ layers. The SiO$_2$/SiN$_x$ layers are used to create a membrane for the OMT probes \cite{Cecil.2020}. But these amorphous dielectric layers are expected to degrade the $1/f$ noise performance of the resonators by introducing an additional source of TLS \cite{Zmuidzinas.2012}. To reduce the presence of TLS in the vicinity of the mm-wave components, we have developed a process for removing the majority of these dielectrics from the wafer. A cartoon cross section of the process is shown in Figure \ref{fig:MembraneFig}. The first step is to lithographically pattern circles slightly larger than the diameter of the OMT probes. The SiN$_x$ layer is etched in a RIE with an optical end point detection system to ensure the etch stops in the SiO$_2$ layer. The resist is stripped and a second lithography step is completed with a slightly larger diameter (roughly 10 ${\mu}$m) than the SiN$_x$ pattern. The SiO$_2$ layer is then etched in 10\% HF buffered oxide etch to remove the SiO$_2$, a common practice for reducing TLS for superconducting circuits. A final lithography step is performed to define the probe and wiring layer. The Nb wiring with a NbN cap layer is deposited and patterned via liftoff. Because material deposited with the sputtering process is not fully line-of-sight, the sidewalls of the membrane step are also coated with Nb providing a continuous film from the top of the membrane layer, over the sidewalls, and on to the bare silicon surface. DC connectivity tests demonstrated electrical connectivity of the Nb film over the step edge. %We have also fabricated resonators using this process as described below.

\begin{figure}
    \centering
    \includegraphics[width=1.0\linewidth]{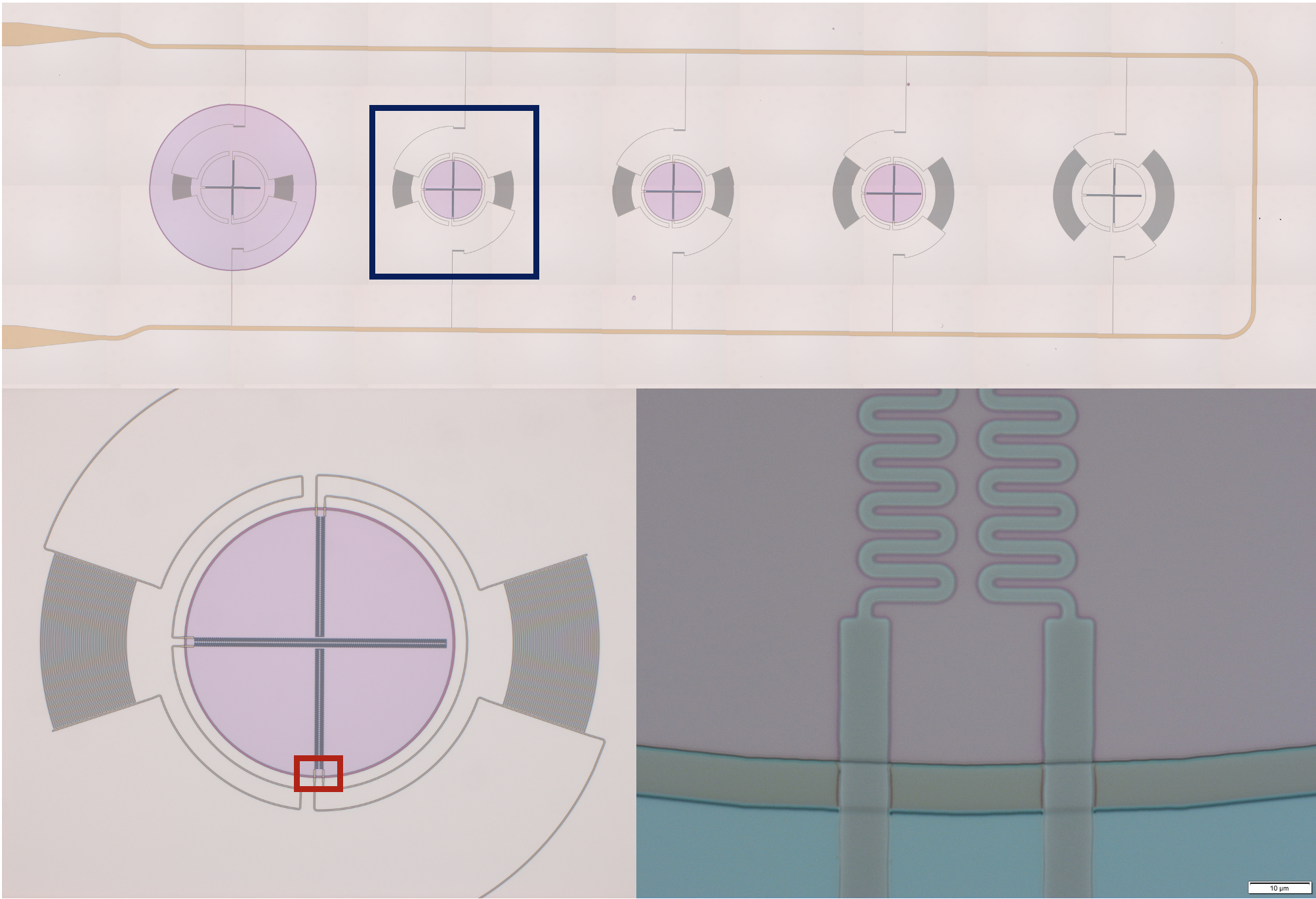}
    \caption{(Top) Photo of a membrane step test chip with three different SPT-3G+ style leKID variations used to characterize the impact of the SiN$_x$ membrane layer: left leKID - entire leKID on SiN$_x$ membrane; center three leKID - inductors on SiN$_x$ membrane and capacitors on bare silicon; and right leKID - entire leKID on bare silicon. (Bottom left) Closeup of leKID in blue box. (Bottom right) Closeup of membrane stepdown shown in red box. The leKID Nb layer transitions from SiN$_x$ to SiO$_2$ to bare silicon, pink to grey to light blue, respectively.}
    \label{fig:StepChipFig}
\end{figure}

\begin{figure}
    \centering
    \includegraphics[width=1.0\linewidth]{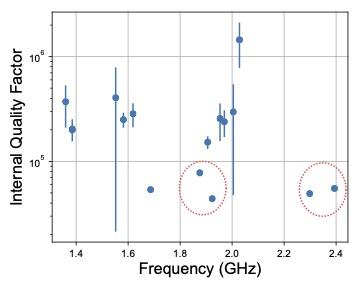}
    \caption{Measured internal quality factors for resonators on a stepdown test chip. Resonators circled with the dashed lines were identified by frequency schedule as those entirely on SiN$_x$. All the resonators fabricated with capacitors on bare silicon, with the exception of one, have internal quality factors greater than $10^5$. Each resonance was measured five times with the median value indicated by a dot and the standard deviation shown with error bars.}
    \label{fig:FT86}
\end{figure}

To fully characterize the stepdown process we have fabricated test arrays using the same well-characterized SPT-3G+ leKIDs described above that now incorporate different amounts of SiN$_x$ under the resonator. Photos of a representative chip are shown in Figure  \ref{fig:StepChipFig}. There are three resonator types: resonators that sit entirely on the SiN$_x$, resonators with inductors on the SiN$_x$ and capacitors on bare silicon, and resonators entirely on bare silicon. Figure \ref{fig:FT86} shows the internal quality factors for a similar chip with sixteen resonators: four entirely on SiN$_x$, eight with inductors on the SiN$_x$ and capacitors on bare silicon, and four entirely on bare silicon. All sixteen resonators are present suggesting that the stepdown does not impact resonator yield. Based upon frequency scheduling we identify the resonators circled by the dashed line as those resonators entirely on SiN$_x$, which also are among the lowest internal quality factors. Eleven of the twelve other resonators have internal quality factors greater than $10^5$; the cause of the 5th resonator with low internal quality factor is unclear. Following quality factor measurements, the next step for this test is to measure the noise performance. 

% The expectation is that the devices entirely on SiN$_x$ will have lower quality factors and/or higher noise than those with capacitors on bare silicon due to the presence of additional TLS sources in the SiN$_x$. The first round of testing with these devices has produced mixed results. In the chip tested cryogenically all resonators were present indicating good fabrication yield and continuity of the step sidewall connection. However, the \textcolor{red}{internal} quality factor of all resonators was lower than expected, $\sim 10^4$, making it difficult to confirm any variation in resonator quality factor based upon the amount of SiN$_x$ under the resonator. We believe the low quality factors are due to an issue with \textcolor{red}{the base vacuum pressure in} the deposition system that deposited the Nb \textcolor{red}{and may have resulted in poor quality Nb}. New device fabrication is ongoing to repeat the test.

%%%%%%%%%%%%%%%%%%%%%%%%%%%%%%%%%%%%%
\section{Prototype Chips}

%Section 4 should cover whatever test results we have to date. \textcolor{red}{Is there anything from the polarization testing on FT47 that we want to include?}.

\begin{figure}
    \centering
    \includegraphics[width=1.0\linewidth]{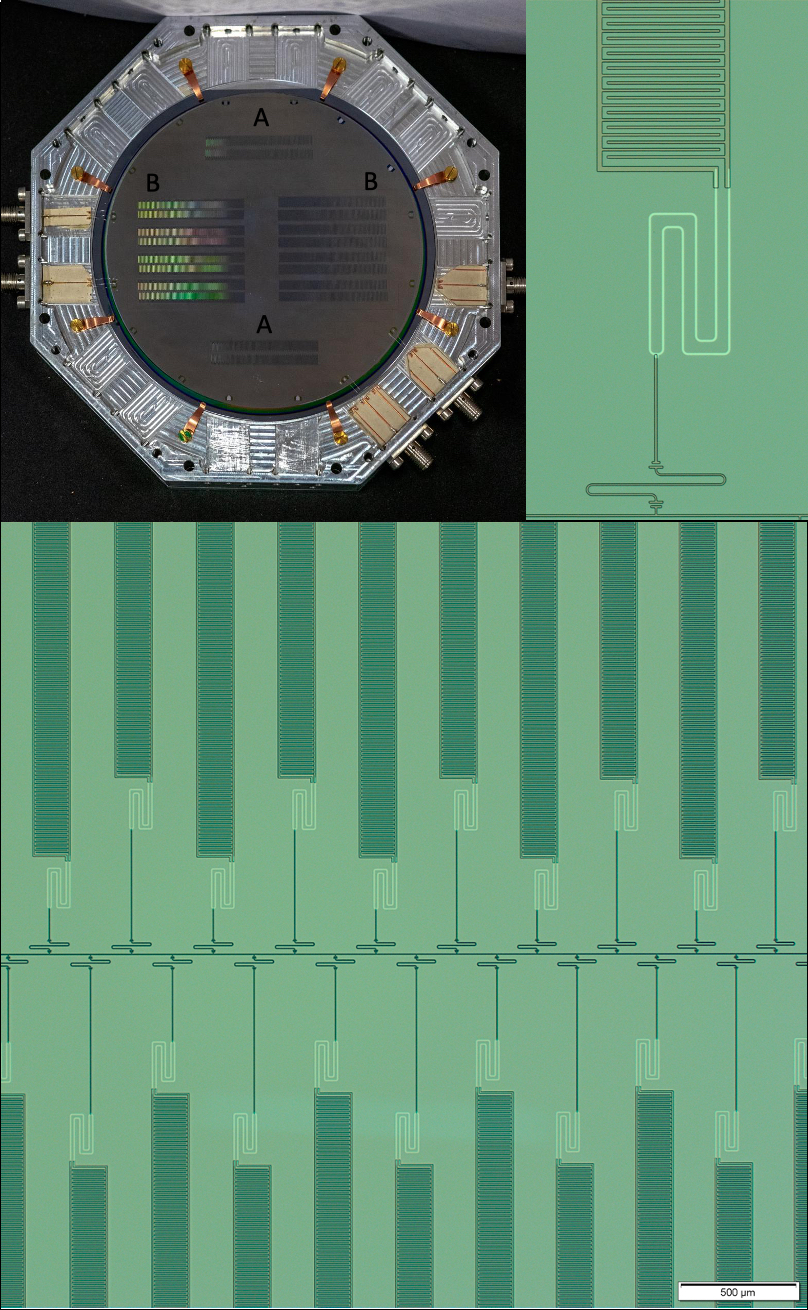}
    \caption{Photos of a multiplexing test wafer designed to evaluate the multiplexing factor of filter-bank spectrometer readout leKIDs. (Upper left) Wafer mounted in sample box. The wafer contains four test spectrometer banks: two (A) with a single, sparsely spaced readout bank and two (B) with four closely-spaced readout banks. (Bottom) Zoom-in on a filter bank showing the array of filters connected to readout leKIDS. (Upper right) Photo of a single channel with channelizing filter at bottom and leKID at top.}
    \label{fig:MuxChipFig}
\end{figure}

Prototype chips to test isolated components of the spectrometer are in %various stages of
development. 
A layout for this chip is shown in Figure \ref{fig:MuxChipFig}. The layout consists of four sets of spectrometer-like devices (optical coupling structures have been left out to simplify the fabrication process). Two of the devices consist of four readout banks, each bank containing 194 leKID channels with the entire device spanning 2.011--2.489 GHz; the 500 kHz resonator spacing is the targeted spacing for the final SLIM instrument. Two of the devices have sparse spacing with 194 leKIDs spanning the entire 500 MHz readout bandwidth. Fabrication of these devices is currently ongoing.

Following testing of the isolated components several steps of integrated testing are required on the path to a deployable focal plane. This testing will be done using a layout that combines a series of test chips and a spectrometer sub-module on a single wafer. Fabrication and testing rounds are needed to verify our resonator trimming process, the optical pass bands, and the microstrip transmission loss at mm-wave for the selected dielectric, which is still to be determined. By combining test chips and sub-modules on a single wafer, once testing verifies a successful fabrication process we will have a potentially deployable portion of the focal plane in hand. Design and layout of these wafers is ongoing.

%%%%%%%%%%%%%%%%%%%%%%%%%%%%%%%%%%%%%%%%%%%%%%%%%%%%
\section{Conclusions}

Line Intensity Mapping is a promising experimental technique for uncovering answers to some of the most fundamental questions about the nature of the universe. Superconducting spectrometers operating in the mm-wave frequency band are well matched to the needs of future mm-wave LIM experiments. We describe the development of superconducting spectrometers designed to operate in the 120--180 GHz band for the upcoming SPT-SLIM instrument which is scheduled to deploy in the 2023--2024 austral summer. This instrument will serve as a pathfinder to demonstrate the viability of ground-based LIM measurements in the mm-wave bands.

%%%%%%%%%%%%%%%%%%%%%%%%%%%%%%%%
% \section*{Acknowledgment}

% \textcolor{red}{Funding acknowledgements go on the front page with author affiliations. Any other acknowledgements can be added here. Otherwise this section will be removed.}

%%%%%%%%%%%%%%%%%%%%%%%%%%%%%%%
% references section
\bibliographystyle{IEEEtran}

\bibliography{references}

\end{document}